# Carbon cluster diagnostics-II: *ExB* velocity filter for large carbon clusters

Shoaib Ahmad[1,2,*], B. Ahmad[1], A. Qayyum[1], M. N. Akhtar[1]
[1]*Accelerator Laboratory, PINSTECH, P.O. Nilore, Islamabad, Pakistan*
[2]*National Centre for Physics, Quaid-i-Azam University Campus, Shahdara Valley, Islamabad, 44000, Pakistan*

Email: sahmad.ncp@gmail.com

## Abstract

An *E×B* velocity filter is described that has been used for the detection and diagnostics of large carbon clusters $C_m$ ($m \leq 10^4$). The velocity and momentum analyses are compared for our special experimental arrangement. We describe the variability of the resolving power of our compact velocity filter as the main advantage over other comparable mass analysis techniques.

## 1. Introduction

We describe a compact, permanent-magnet-based *E×B* Wien velocity filter that has been developed for the detection and diagnostics of carbon clusters. Carbon clusters $C_m$ ($m \leq 10^4$) can be produced and emitted from the cluster-forming devices utilizing processes like arc discharge [1], laser ablation [2,3] and hollow cathodes [4]. Different mass analyzing techniques have been employed for the detection of such a large range of cluster masses. These include time-of-flight (TOF), momentum analysis and velocity filtration employed in the cluster-forming techniques [1-4]. All of these cluster diagnostic techniques have certain advantages as well as specific limitations. We are interested in on-line analysis of clusters for use as continuous beams. Therefore, velocity analysis has certain ad-vantages over the momentum analysis and we shall describe the two techniques in detail in the next sections. The TOF technique cannot be used due to its pulsed nature although it has superior resolution especially in the higher mass range.

The main emphasis during the design and fabrication stages has been on keeping the overall size of the velocity filter within manageable dimensions. The neodymium iron boron bar magnets of 14 mm]14 mm cross section and 45 mm length are the centre pieces of the compact filter design. A Mild Steel return core of 100 mm diameter and 60 mm length houses the entire filter including the compensating electric field plates. The multiple units can be coupled together to (a) increase the resolution, (b) correct the ellipticity of the resolved lines and (c) increase the collimation of the resolved species. The total weight of a twin-unit filter is (1 kg and the only control element is a $500 V DC supply. The momentum analyzer of comparable resolution weighs more than 200 kg and is much less manoeuvrable.

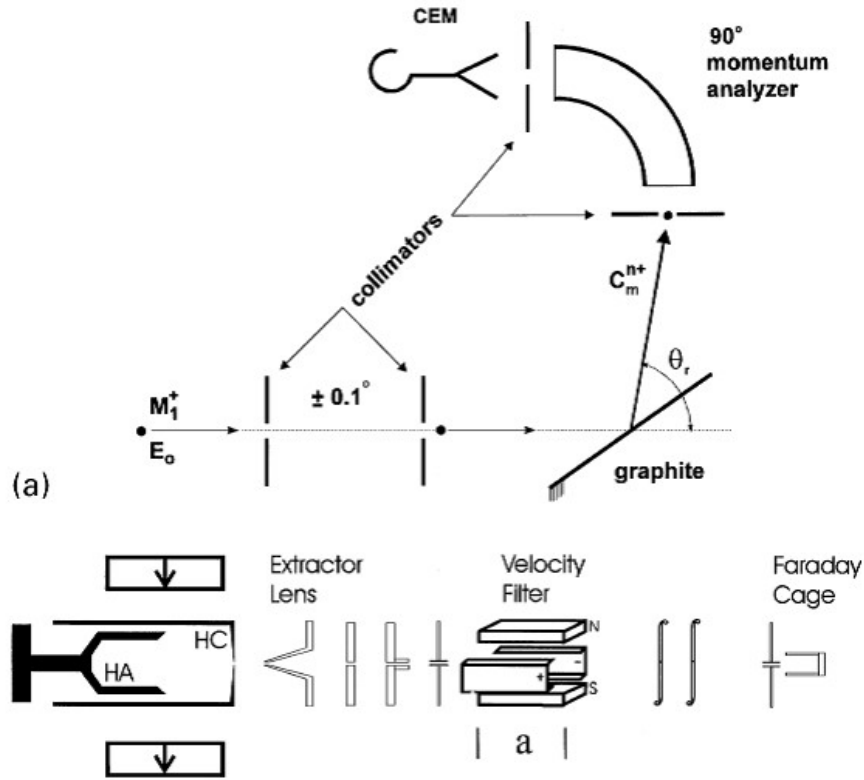

Fig. 1. The two experimental setups are shown in 1(a) and 1(b). (1a) has the Direct Recoil detection with heavy masses of $M_1^+$, energy $E_0$ and the clusters $C_m^{n+}$ recoiling at angle $\theta_r$, being analyzed with a 90-degree momentum analyzer. 1(b) shows the cluster ion source, extraction lens, collimators, the velocity filter of dimension $a$. The Faraday cage is at distance $l$ from the filter. The source is composed of a Hollow Cathode (HC), Hollow Anode (HA) and a set of hexapole-bar magnets shown with arrows.

## 2. Experimental procedure

We describe two sets of experimental arrangements that have been used in our laboratory for the detection and diagnostic of carbon clusters. In the first setup we momentum analyze the Direct Recoiling Clusters (DRCs) under heavy-ion irradiation of graphite surface [5] and in the other the velocity analysis of the clusters emitted from a graphite hollow cathode ion source [3] is presented. The two diagnostics are compared for large carbon cluster identification but with largely different cluster energies. The momentum analysis is performed on DRCs that have energies $\leq$ 200 eV for a 100 keV $Kr^+$ beam. Whereas, the energy of the clusters from the hollow cathode source is determined by the extraction voltage Vext and can have any value but the practical limits are $V_{ext} \leq$ 5 keV as will be discussed later.

For the momentum analysis we use a 250 keV heavy-ion beam facility where $Ar^+$, $Kr^+$ and $Xe^+$ beams of >1 mm diameter and energy between 50 and 250 keV can be delivered to a target 2 m from the end of the accelerator tube. The facility is equipped with the duoplasmatron operating at pressure $10^{-2}-10^{-3}$ mbar. The experimental setup is shown in Fig. 1a where the beam as well as the recoil particles' collimators with $\leq \pm 0.1°$ divergence are shown along with a 90° momentum analyzer. Experiments are performed with the target chamber at pressures $10^{-7}$ mbar maintained with an ion pump. A Channel Electron Multiplier (CEM) is used for cluster detection. Momentum analysis of clusters is desirable to unambiguously characterize the *m/q* values but the required magnetic fields become unrealistically large for experimental arrangements like ours. For example, in the case of $\theta_{DR} = 79.5°$, a large magnet is needed with $B_0\rho \geq 4T$m for resolving $C_{60}$. For $\theta_{DR} = 87.8°$ we performed momentum analysis with $B_0\rho \geq 0.06\,Tm$. Such an analyzer is appropriate only for smaller recoil angles.

The permanent-magnet-based *E×B* velocity filter can perform mass analysis in a characteristic way. All masses are deflected by the fixed magnetic field according to their respective masses. The straight-through beam contains the desired mass at the compensating electric field $\varepsilon_0$. In principle, a velocity spectrum always contains all masses irrespective of their mass: the resolvability on the other hand is dependent upon certain design features. Fig. 1b shows the experimental arrangement for the detection of carbon clusters $C_m$ from a cusp field, hollow cathode ion source [4]. A well-collimated set of extraction lens setup provides a $\pm 0.1°$ beam to the velocity filter of effective length *a*. A picoamp meter measures the analyzed masses on a Faraday cage *l* mm away from the exit of the filter. We have made measurements with *l* =1500 mm, though in principle, longer flight paths increase the dispersion of the selected species.

## 3. Results

Fig. 2 is the mass spectrum of the momentum-analyzed cluster species that are directly recoiling from a graphite surface under bombardment by 100 keV Kr` beam at recoil angle $\theta_{DR} = 87.8°$ taken from Ref. [5]. The clusters of constitution $C_m$ receive recoil energy $E_r$ given by Bohr [6] as

$E_r = \{(4m_1m_2)/(m_1+m_2)^2\}E_0 \cos\theta_{DR}$.

The respective masses are analyzed as a function of the resolving magnetic field $B_0$. The sputtered species can also be seen in the spectrum as these have energies 410 eV. The entire range of the clusters from the very small to the heavy ones are present in the figure. The two essential features of any momentum analysis are (a) the presence of the multiply charged ions and (b) the dependence of the highest mass on the ultimately available magnetic field $B_0$. Both of these features can be clearly identified from this spectrum.

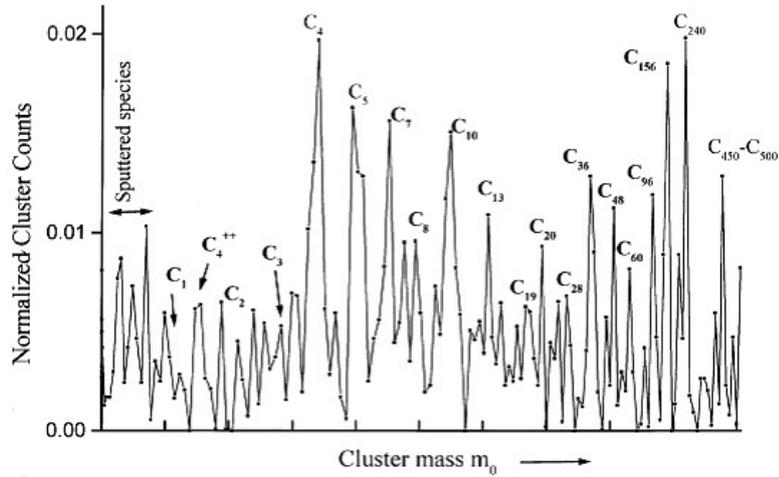

Fig. 2. The momentum spectrum of the Direct Recoiling clusters from graphite surface irradiated with 100 keV $Kr^+$. The low-energy sputtered species are also present as are the multiply-charged ones.

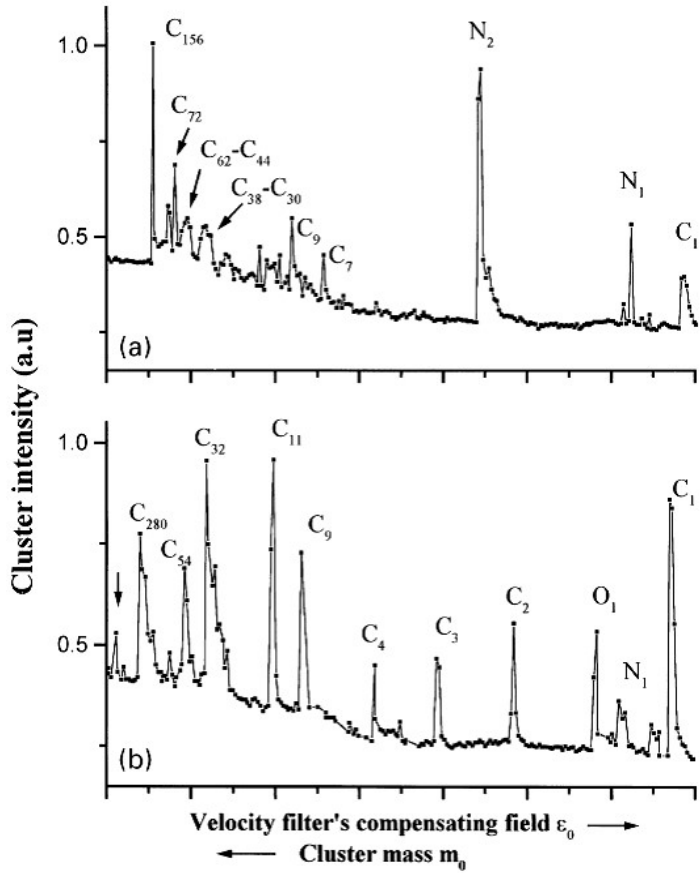

Fig. 3. Velocity spectra from the source at $V_d$=0.85 kV, $I_d$=100 mA and $V_{ext}$=1 kV. (a) The spectrum shows the entire range of clusters. (b) The same source parameters but spectrum taken after 60 min operation.

Figs. 3(a) and b are the velocity spectra of the clusters extracted from the carbon cluster source. The two consecutive spectra show the clustering mechanisms going on in the regenerative sooting plasma [7] of the source. Fig. 3a has a small constituent of the $N_2$ gas but the spectra shows large peaks due to $N_2$ and $N_1$ as well as contributions from higher carbon clusters from $C_7$ to $C_{156}$. The same species is greatly reduced in intensity in Fig. 3b that was taken 60 min afterwards. The $C_1$ peak is also enhanced as well as the heavier carbon clusters right up to $C_{280}$ and even a small peak for $C_{3100}$ identified with the arrow is shown in the figure. The ratio of $C_1$, $N_1$ and $N_2$ to clusters $C_m$ ($m \geq 2$) is 1 : 2 in Fig. 3a and rises to 4 : 1 in Fig. 3b. We consider the support gas ions as well as the monatomic species $C_1$ as the main sputtering agent that releases carbon atoms from the cathode and act as the source of sooting of the plasma. The mechanisms of cluster formation and the evolving phases of the sooting plasma under different experimental conditions have been discussed elsewhere [7].

## 4. Resolution of $E \times B$ velocity filter

An $E \times B$ velocity filter is generally designed with a desired mass range in mind. Since higher compensating electrostatic fields are required for lighter particles for any fixed magnetic field $B_0$, the limit is set on the one hand by the lightest mass with the highest velocity $v_{max}$ and the highest detectable mass with the minimum velocity $v_{min}$ on the other. The requirement for the detection of heavy carbon clusters imposes the condition of achieving the highest possible magnetic field $B_0$ and sets other parameters accordingly. This was done with our velocity filter. We have $B_0 = 0.35$ T on the axis of the filter between the poles that are 10 mm apart. A set of four permanent bar magnets of cross section 14 mm×14 mm and 45 mm length are stacked in sets of two each on the inside of a cylindrical soft iron core. The spacing between the North and South poles is set by two C-shaped insulators. Inside surfaces of these insulators contain the specially shaped conductors that provide the balancing electric field $e_0$ for the straight-through masses. These electrodes are slightly extended outwards to compensate for the magnets' edge effects.

The resolution of the $E \times B$ velocity filter is determined by the dispersion $d$ of masses $m_0 \pm \delta m_0$ from the resolved mass $m_0$ that travels straight through the filter with velocity $v_0$ (=$B_0/\varepsilon_0$). It is given as $d \propto al(\delta m_0/m_0)(\varepsilon_0/V_{ext})$ where $a$ and $l$ are the lengths of the velocity filter unit and the flight path, respectively [8]. These two-dimensional parameters are shown in Fig. 1b. For a given ratio $\delta m_0/m_0$, the dispersion $d$ can be enhanced by (1) stacking multiple filter units since $d \propto na$, $n$ being the number of filter units, (2) increasing the flight path $l$ and (3) enhancing $V_{ext}$. The multiple unit filter has the axes of mass resolution of successive units rotated by 90° so that the beam ellipticity introduced along the "led axes is compensated by each subsequent unit. Therefore, stacking in sets of two is desirable. The flight path $l$ cannot be in-creased beyond an optimum length as the beam divergence can over-compensate the dispersion achieved by the elongation of $l$. We have found $l$=1500 mm as an optimum for our experimental conditions. Increasing $V_{ext}$ implies a larger spread between the heavier clusters at the lower velocity end of the spectra. This is a desirable feature but caution should be exercised when applying too large extraction fields since these introduce multiple ionization and can distort an already complicated cluster spectrum. We have used a maximum of $V_{ext}=$ 4 kV.

## 5. Higher mass resolution

As discussed earlier, the resolution of a velocity filter varies throughout the spectrum. This is due to the uniform step size of the electrostatic potential applied across the compensating field plates $\varepsilon_0$. Therefore, a difference of $\pm 1$ V may resolve 1 amu in the high velocity (i.e., low mass) part of the spectrum while the same field might just be able to resolve $\pm 100$ amu when analyzing the largest masses. This is the main constraint against the use of velocity filters for mass analysis with constant electric field step $\varepsilon_0$. However, as mentioned in Section 4, the presence of even the heaviest masses is always indicated in the low-velocity part of the spectrum. The enhancement of resolution could be achieved for such masses by reducing the $\varepsilon_0$ steps for a given $V_{ext}$ and then increase the extraction voltage in stages.

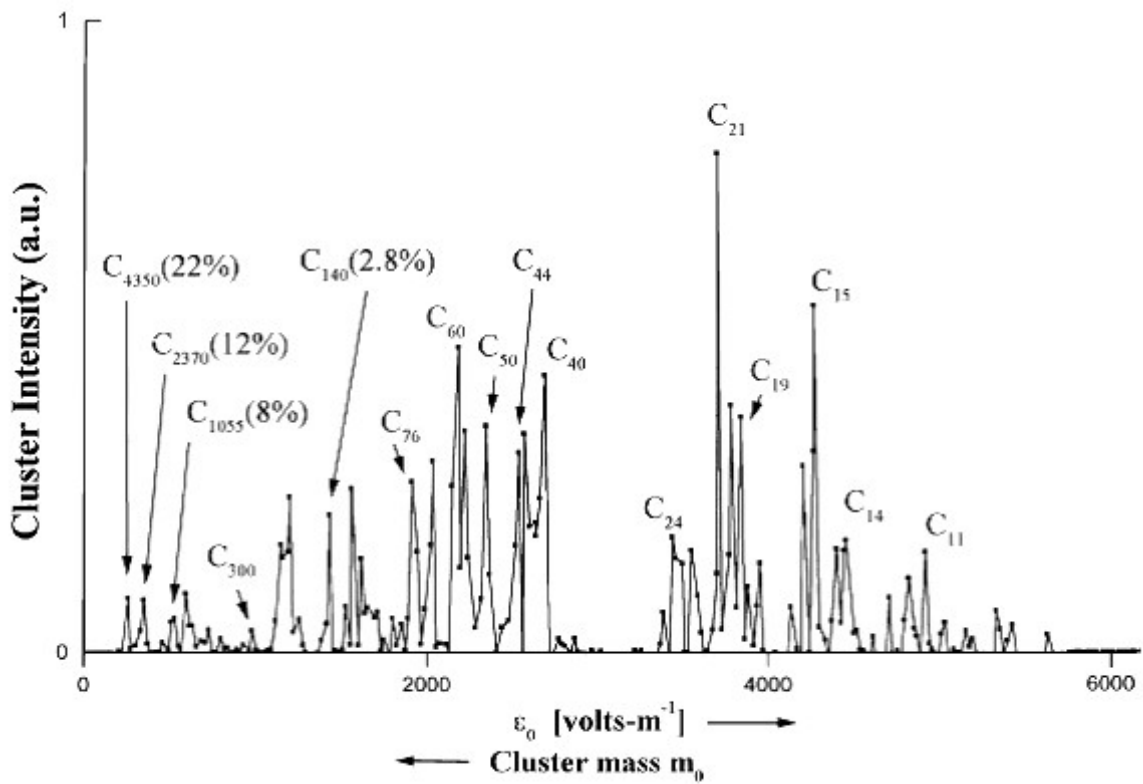

Fig. 4. The spectrum from the well-sooted source operated at $V_d=0.5$ kV, $I_d=50$ mA and $V_{ext}=2$ kV with Ne at $\approx$ 60 W for 20 hours.

Fig. 4 is the velocity spectrum from the well-sooted cluster source that was operated with Ne at 60 W for 20 h. The figure shows the masses higher than $C_{10}$ up to $C_{4350}$. The cluster mass $m_0$ is inversely proportional to the resolving field $\varepsilon_0$ as can be seen from the oppositely directed arrows on the x-axis.

On the low-velocity side of the spectrum we have the heaviest resolved peak around $C_{4350}$ with asymmetrical mass fluctuations $\pm^{980}_{310}$. In the figure the maximum error is shown as 22% while the average error is 11%. It can be seen that the larger error around mass $C_{4350}$ (i.e., 980 amu) is due to higher mass per unit electric field for the lower velocity side of the peak. The second leak is for $C_{2370}$ with error $\pm^{290}_{225}$ with 12% maximum error as opposed to average of 10.6%. Similarly, for the third peak at $C_{1055}$ with mass determination error $\pm^{89}_{73}$ has the maximum and the average error reduced to 8%. For the next peak at $C_{300}$ the error is reduced to < 3%. We then have the better resolved peaks at $C_{76}$, $C_{60}$, $C_{50}$, $C_{44}$ and $C_{40}$ all bundled together. The remaining unmarked peaks are all with a mass difference of $C_2$, for example, $C_{58}$ and $C_{42}$ are next to $C_{60}$ and $C_{44}$, respectively. Up to $C_{40}$ we have the usual fullerene spectrum that leads to the cluster series starting from $C_{24}$, $C_{21}$, $C_{19}$, $C_{15}$ to $C_{11}$. In between these peaks are the remaining peaks with a difference of one C1 unit. This is a familiar pattern of carbon cluster fragmentation for the fullerenes $C_m(m \geq 32)$ and rings and linear chains $C_m(m \leq 24)$ [2,3].

## 6. A comparative assessment of the momentum, velocity and time-of-flight spectrometry

We sum up the preceding discussion on the momentum and velocity spectroscopy with a comparative assessment of the three most widely used techniques, namely, the momentum, velocity and the time-of-flight (TOF). In Table 1 we have com-pared their characteristic properties. The first and the most crucial property is the limit on the highest analyzable mass. In case of momentum analysis, the limit is set by the maximum available field $B_0^{max}$, the other two techniques do not suffer from this handicap. There are, however, inherent limitations on the resolution in case of velocity analysis and it has been thoroughly discussed earlier. The mass resolution is variable except for the magnet which is fixed. Magnet being heavy is also less manoeuvrable compared with the other two competitors and is bulky and heavy. In terms of the economic viability the velocity filter is the cheapest since the cost of permanent magnets and a dual-voltage power supply in our case ≈ $2000. Therefore, in our conclusion, a compact velocity filter with variable resolution is a versatile device for the mass analysis of even fairly heavy ions

Table 1
A comparison of the three cluster detection techniques is shown for six characteristic properties

| Property | Device | | |
|---|---|---|---|
| | $E \times B$ | Magnet | TOF |
| Limit on highest mass | All masses | Depends on max field $B_0$ | All masses |
| Resolution | Variable | Fixed | Variable |
| Maneouverability | Light and compact | Heavy and fixed | Light |
| Dimensions | 10 cm × 20 cm dia | Depends on $\rho$, ~ meters | Flight tube |
| Weight | ≤ 2 kg per unit | ~ 100–200 kg | Few kg |
| Price | Magnets, PS ~ 2 k$ | Magnet, PS ~ 30 k$ | Tubes, PS ~ 10–20 k$ |

# 7. Conclusions

In this communication, we have presented results from a compact velocity filter that was originally designed to identify large carbon clusters $C_m$ ($m \leq 10^4$) emitted from sooting plasma. The velocity filter's performance has been compared with a momentum analyzer built in our laboratory also for the analysis of carbon clusters. We have seen that whereas the momentum analyzer can be employed for the same purpose, it has inherent limitations on the highest masses that it can resolve. We have also compared the overall mass resolving capabilities of the three devices including the other widely used time-of-flight (TOF) technique. We have shown that although all three may have their specific uses, the velocity filter can compete with TOF and is certainly more cost-effective and manoeuvrable as opposed to the magnetic analyzer. The velocity filter is the only straight through device and one can improve upon its resolution by stacking multiple units; a feature that is not matched by any other device.